# Structural Optimisation: Biomechanics of the Femur

Dr Andrew T. M. Phillips

Imperial College London, Structural Biomechanics,
Department of Civil and Environmental Engineering, Skempton Building,
South Kensington Campus, London SW7 2AZ

www.imperial.ac.uk/structuralbiomechanics

**Abstract:**

A preliminary iterative 3D meso-scale structural model of the femur was developed, in which bar and shell elements were used to represent trabecular and cortical bone respectively. The cross-sectional areas of the bar elements and the thickness values of the shell elements were adjusted over successive iterations of the model based on a target strain stimulus, resulting in an optimised construct. The predicted trabecular architecture, and cortical thickness distribution showed good agreement with clinical observations, based on the application of a single leg stance load case during gait. The benefit of using a meso-scale structural approach in comparison to micro or macro-scale continuum approaches to predictive bone modelling was achievement of the symbiotic goals of computational efficiency and structural description of the femur.

**1. Introduction:**

It has long been observed that the structural composition of the femur (thigh bone), is adapted in response to the mechanical environment that it is subjected to. In cross-section the proximal femur (close to the hip joint), is composed of two distinct types of bone as illustrated in Figure 1, which shows a labelled diagram of a coronal (frontal) slice of the proximal femur. Cortical bone is formed from a layer of low porosity, high stiffness bone, of varying thickness on the outside of the femur. Trabecular bone is formed from a series of struts, giving rise to a structure in which there is a spacial variation of continuum level porosity and directionally dependent stiffness throughout the femur. At a tissue level cortical and trabecular bone can be considered to be the same material, with varying material properties being a result of the architecture. Both the varying thickness of the cortical bone and the structural properties of the trabecular bone are thought to be a result of the forces placed on the femur, which include the joint contact forces at the hip and knee joints, and muscle forces, which act on the cortex (cortical surface) of the femur to facilitate balance and movement. It is generally accepted that the resulting structure of the femur is optimised to withstand the applied forces using a minimal amount of material.

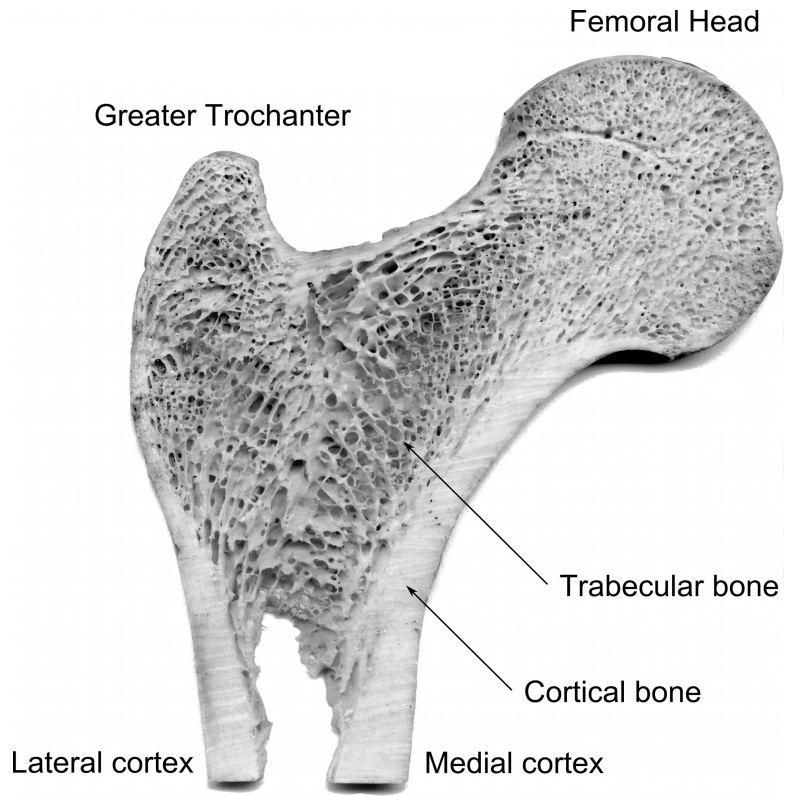

Figure 1: Labelled diagram of a coronal slice from the proximal femur

It has been hypothesised that the structure of trabecular bone in particular follows trajectories of compressive and tensile stress, resulting in an optimised structure. This hypothesis originates from observations by Culmann (an engineer) and von Meyer (an anatomist) that the internal structure of a frontally sectioned proximal femur resembled the sketched stress trajectories of a curved (Fairbairn) crane (Culmann 1866, von Meyer 1867). Culmann was a pioneer of graphical methods in engineering, publishing *Die graphische Statik* (*Graphical Statics*) in 1866. It is of note that in drawing the trajectory arrangement of trabeculae in the proximal femoral head, von Meyer did not imply that the trajectories were orthogonal. In his work *Das Gesetz der Transformation der Knochen* (*The Law of Bone Remodelling*) (Wolff 1986) Wolff criticised von Meyer for this, and produced a trajectory diagram of the proximal femur in which trajectories met at right angles, the implication at a continuum level being that trajectories occur along lines of principal stress. Koch, in his work *The Laws of Bone Architecture* (Koch 1917) also argues that trabecular trajectories should be orthogonal, based on trajectories following principal stress directions at a continuum level. The reader is directed to Skedros & Baucom (2007) for an in-depth discussion of this assumption. Suffice it to say that under examination trabecular architecture is demonstrated not to be orthotropic throughout the proximal femur. It has been suggested that non-orthogonal trabecular intersections offer increased shear resistance (Pidaparti & Turner 1997), significant if the femoral structure is considered to be optimised for a range of loading scenarios. Wolff's observations are often now erroneously referred to as 'Wolff's law' in studies investigating bone modelling and remodelling (resorption and apposition of bone). Figure 2 shows diagrams of the representations of the cross-sectional trabecular arrangement in the proximal femur as proposed by von Meyer, Wolff and Koch.

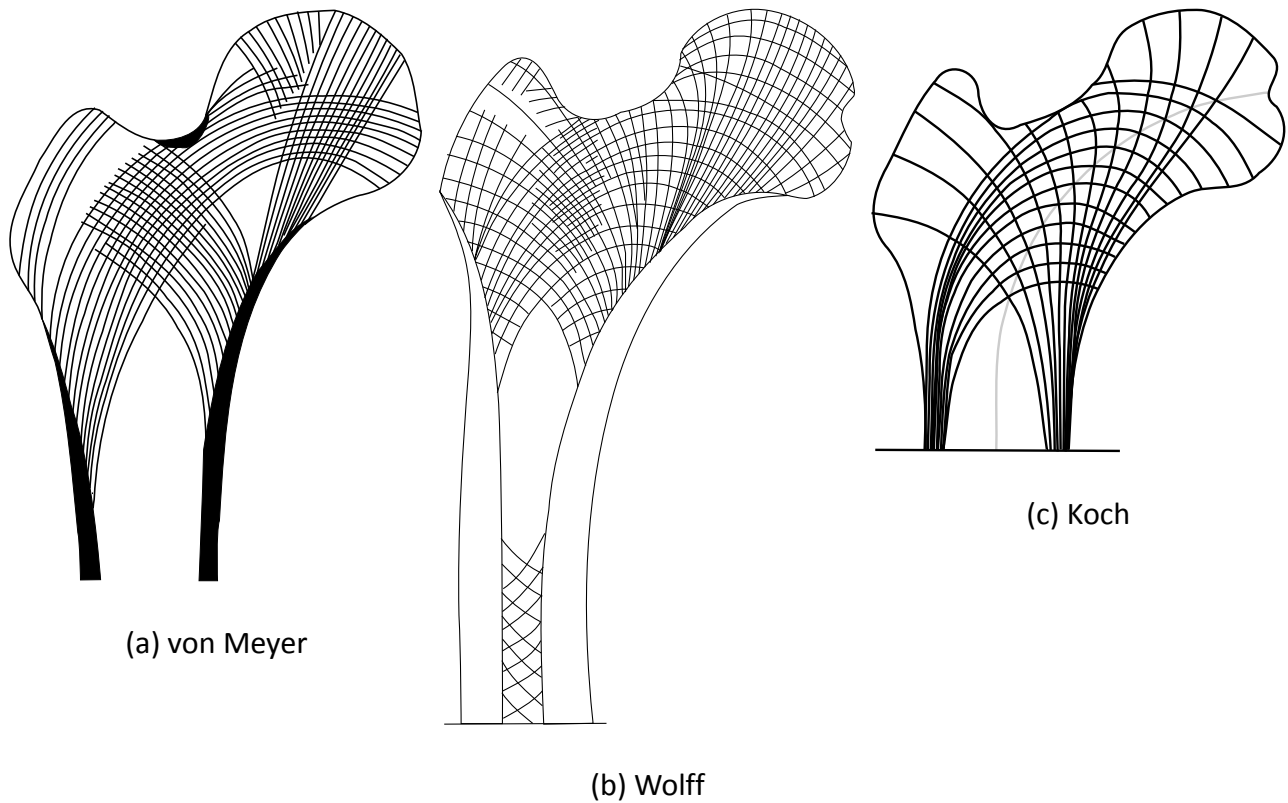

Figure 2: Representations of the trabecular arrangement in the proximal femur based on (a) von Meyer (1867) , (b) Wolff (1986) and (c) Koch (1917).

Singh et al. (1970) along with others identify five distinct trabeculae groups in the proximal femur. These are described briefly: the principal compressive group carries load from the hip joint, through the femoral head towards the medial (close to the centreline of the body) cortex; the primary tensile group arches from the lateral (distant to the centreline of the body) cortex through to the femoral head; the secondary compressive group spans in a diffuse manner from the medial cortex across the femoral shaft; the secondary tensile group spans in a diffuse manner from the lateral cortex across the medial shaft; the greater trochanter group is thought to be a tensile group running within the greater trochanter. Also noted are the presence of Ward's triangle, an area of low trabecular density, as well as the femoral canal running between the medial and lateral cortex of the femoral shaft.

Frost (2003) introduced the idea of the 'mechanostat'. The essence of the mechanostat is the notion that bone adapts towards a target strain, similar to a thermostat regulating heating and cooling towards a target temperature. Based on the mechanostat if bone experiences a strain higher than the target strain apposition occurs, causing the strain to revert towards the target strain. While if bone experiences a strain lower than the target strain resorption occurs, again causing the strain to revert towards the target strain. Frost also proposed a lazy zone for the mechanostat, similar to allowing a thermostat to maintain temperature between two values, only regulating heating or cooling when the temperature goes outside the range. In addition Frost suggested values for complete bone resorption as well as yield and fracture strain.

Although it is clear that a significant proportion of the biomechanical behaviour of the femur can be attributed to its structure, it has not been common for a structural approach to be taken when attempting to resolve the distribution of cortical and trabecular bone within the femur, and in particular the varying thickness of the cortical bone, and varying structural properties of the trabecular bone. Rather is has been common to adopt a continuum level approach. Continuum level approaches can be split into two categories; those in which the solid elements used to represent bone in the femur are larger than the individual structural elements in the femur, referred to as a macro-scale approach; those in which the solid elements used are smaller than the individual structural elements, referred to as a micro-scale approach. Macro-scale predictive models treat the femur as a complete continuum without voids, varying material properties for each element in an attempt to replicate the overall behaviour (Carter et al. 1989). Material properties used in subject specific macro-scale models of the femur are derived based on CT (computed tomography) data, where the greyscale values can be converted to an isotropic value for Young's modulus (Bitsakos et al. 2005). Micro-scale predictive models treat the femur as a binary system with bone either being present or not, and with a single material property being adopted for bone (Tsubota et al. 2009). The arrangement of solid elements in subject specific micro-scale models are derived based on µCT (micro CT), using a single greyscale threshold value to distinguish between solid bone material and void (van Rietbergen et al. 2003, Verhulp et al. 2006). Macro-scale modelling has the advantage of being computationally efficient, but suffers from treating trabecular bone in particular as an isotropic material, thus not representing the structure of the bone. Micro-scale modelling suffers from being extremely computationally demanding, and is sensitive to the threshold value chosen to delineate bone. In addition µCT imaging is time consuming and can only be carried out on cadaveric samples due to the associated radiation dose.

Macro-scale 2D (Miller et al. 2002) and 3D (Geraldes & Phillips 2010) orthotropic predictive models have been introduced recently in an attempt to provide additional directional information in comparison to isotropic predictive modelling (Carter et al. 1989). These are based on orientating material properties with respect to principal stresses, and altering material properties based on associated local strains. Anisotropic material models may offer more directional information, although it is conceptually challenging to imagine what stimuli could be used to drive such models.

Thus it is is desirable to investigate a method of capturing the structural behaviour of the femur in a way that has not been achieved using a macro-scale model, but which is more computationally efficient than using a micro-scale model. This study presents an iterative structural model of the femur, in which shell elements are used to represent cortical bone, and beam elements are used to represent trabecular bone. The model can be viewed as operating on the meso-scale, in that the individual structural elements are larger than those found *in-vivo*, but are believed to be capable of capturing the overall structural behaviour of the femur.

**2. Method:**

The initial structural model of the femur was created based on a CT scan of a *Sawbones* fourth generation composite femur (#3403). The CT data was processed in *Mimics* to create a mesh composed of 113103 four noded tetrahedral elements of the internal volume of the femur with an average element edge length of 4.50mm (SD 1.23mm). The node and element information for the mesh was then exported and adapted using *MATLAB*.

**2.1 Base model:**

The nodes and element faces on the surface of the continuum tetrahedral mesh were used to define three noded triangular shell elements, taken to be representative of cortical bone. An initial thickness of 0.1mm was assigned to all shell elements. The internal nodes of the continuum tetrahedral mesh were used to develop a network of truss elements. This was done by finding the nearest 16 nodes to each node, and defining two noded truss elements connecting between the node under consideration and its 16 nearest neighbours. Truss elements were assumed to have a circular cross-section with an initial radius of 0.1mm. Through this process an initial structural model of the femur was created with 10410 cortical shell elements, and 218717 trabecular strut elements. Using a minimum connectivity at each node of 16, it was assumed that a range of trabecular strut directionalities would be present, allowing specific regional directionalities to develop during the iterative bone modelling process. Further work will investigate changing the initial number of connections. It should be noted that choosing the 16 nearest neighbours to each node, results in a minimum connectivity of 16, while the maximum connectivity is higher. The material assigned to all elements was bone, with a Young's modulus of 18,000N/mm², and a Poisson's ratio of 0.3, based on literature values for bone at the tissue level (Turner et al. 1999).

The base model, and each successive iteration was subject to a combination of loads acting on the proximal femur representative of the joint contact force, the action of the abductor muscles and the iliotibial band at the point of maximum hip joint contact force associated with normal walking (Bergmann et al. 2001). Fixed boundary conditions were applied on the condyles on the distal femur (close to the knee joint). This load case was adopted as it represents the maximum load combination for the most frequent daily activity (Morlock et al. 2001). For this preliminary study loads were applied as point loads as shown in Figure 3, with the exception of the joint contact force which was distributed over seven nodes. This was considered reasonable as the insertion areas of the abductor muscles and iliotibial band on the femur are relatively small, in comparison to the large surface area of the femoral head, through which loads are transferred at the hip joint. Values of the loads in the adopted spacial coordinate system are given in Table 1.

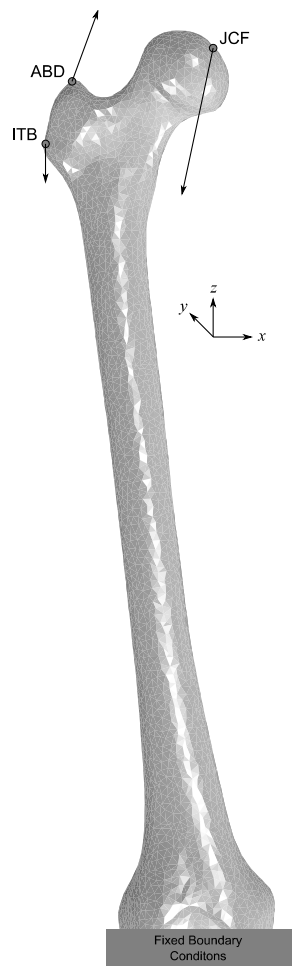

Figure 3: Finite element model of the femur showing boundary conditions and loading. JCF is the joint contact force, ABD is the abductor muscle force, and ITB is the iliotibial band force.

Table 1: Loading applied to the finite element model

| Force | x-axis component (N) | z-axis component (N) |
|---|---|---|
| Joint contact force | -520 | -2445 |
| Abductor muscles | 428 | 1175 |
| Iliotibial band | 0 | -625 |

## 2.2 Iterative Process:

Based on the Mechanostat proposal (Frost 2003) successive iterations of the base model were run with the thickness of individual cortical bone shell elements, and radii of individual trabecular bone truss elements being varied with each iteration according to the resulting strain environment. The iterative process was controlled using *MATLAB*, while successive finite element models were run using the *Abaqus/Standard* solver. The adopted values of absolute strain associated with the dead zone, bone resorption, the lazy zone and bone apposition, as well as target strain are given in Figure 4. For reference bone yield and fracture strain values are indicated, although it should be noted that these were not considered during this study.

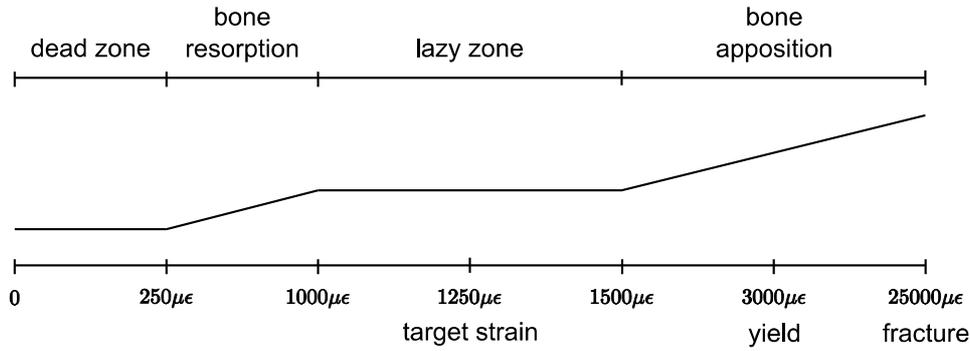

Figure 4: Adopted values of strain associated with the dead zone, bone resorption, the lazy zone, and bone apposition, as well as the target strain.

The cross-sectional area of each of the truss elements with an absolute axial strain value outside of those defined by the lazy zone was adjusted linearly between successive iterations based on the algorithm:

$$A_{i+1} = A_i \frac{\varepsilon_i}{\varepsilon_t} \quad \text{(Equation 1)}$$

where $A_i$ is the cross-sectional area for the current iteration, $\varepsilon_i$ is the absolute axial strain, $\varepsilon_t$ is the target absolute axial strain, and $A_{i+1}$ is the new cross-sectional area.

The cross-sectional areas of the truss elements were allowed to vary between values associated with minimum and maximum radii of 0.1mm and 2.0mm. Although somewhat arbitrarily chosen these values were considered to correlate on a meso-scale with observed values for trabecular thickness on a micro-scale.

Based on the mechanostat resorption limit truss elements with a minimum radius of 0.1mm and an axial strain below 250με in one iteration were assigned a near zero radius of 0.001mm for the next iteration. This allowed for the effective removal of truss elements from the model, while maintaining the numerical stability of the computational model. In subsequent iterations if the strain in one of these near zero area truss elements rose above 2,500,000µε, in the next iteration the element was assigned the minimum radius of 0.1mm, in effect allowing removed elements to regenerate. The value of 2,500,000µε was chosen based on the ratio of the minimum cross-sectional area and the near zero cross-sectional area.

The cross-sectional thickness of each of the shell elements with a maximum absolute principal strain value, occurring on either the top or bottom surface of the the shell element, outside of those defined by the lazy zone was adjusted linearly between successive iterations, in a similar manner to the cross-sectional areas of the trust elements.

Although it would be possible to isolate bending and axial strains within the shell elements, and produce a more sophisticated algorithm on this basis, a linear approach was found to provide a suitable convergence rate. The thickness values of the shell elements were allowed to vary between minimum and maximum values of 0.1mm and 10mm. The values correlate with observed values in the femur (Stephenson & Seedhom 1998, Treece et al. 2010). The mechanostat resorption limit was not applied when modelling the cortical bone.

In order to improve the computational efficiency in *Abaqus* instead of assigning each trabecular and cortical element a unique radius or thickness, elements were classified into 1 of 256 sections, spaced equally between the limits for radius and thickness as appropriate. The model was considered to have reached a converged state when 99% of the trabecular and cortical elements remained in the same section classification between consecutive iterations, and when the change in the effective number of trabecular elements (non near zero area elements) was below 0.1% between iterations. The developed iterative model was found to be robust, with numerical stability maintained when input parameters were varied.

**Results:**

Convergence was found to take 42 iterations based on the chosen convergence criteria. The effective number of trabecular elements in the converged model was 63,682. The total volumes of trabecular and cortical bone were found to be 16,508mm$^3$ and 81,506mm$^3$ respectively. The mean and standard deviations of the nodal connectivity for the trabecular bone was found to change from 21.30 (SD 5.51) for the base model, to 9.96 (SD 5.02) for the converged model.

Figure 5 shows diagrammatic plots of a 5mm thick slice of the derived structure of the proximal femur at the first, fifth, and converged iterations of the model. Also shown is the original model configuration. In order to enhance the appearance of the trabecular structure, the active trabecular elements are scaled based on the cross-sectional area, while the cortical bone is scaled based on the thickness. The greyscale value of the trabecular bone is also varied based on the cross-sectional area, with darker areas representing larger cross-sectional areas, while cortical bone is shown in grey. For cortical bone, element edges, as opposed to element faces, have been plotted so as not to give a false impression of thickness over the depth of the slice.

Figure 6 shows shows diagrammatic plots of a 5mm thick slice of the proximal femur in the converged state highlighting parts of the structure predominantly under tension or compression, based on the applied loading and boundary conditions.

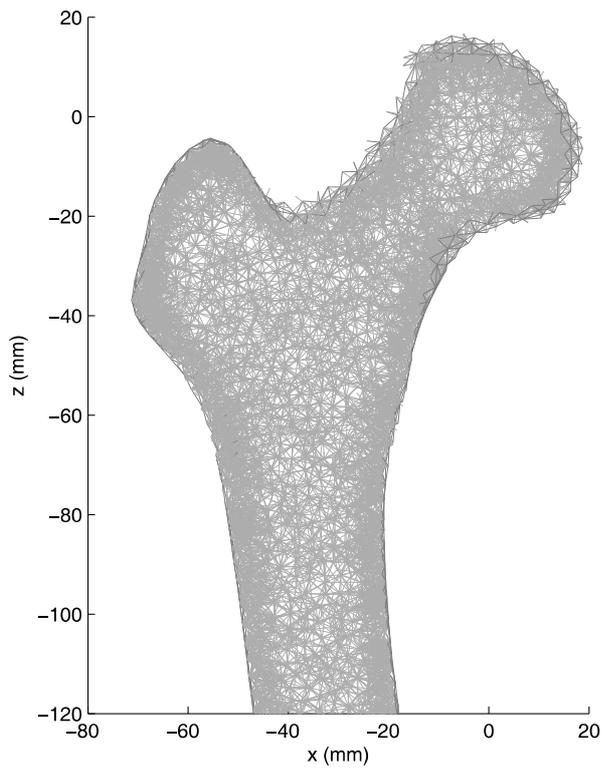
(a) Original configuration

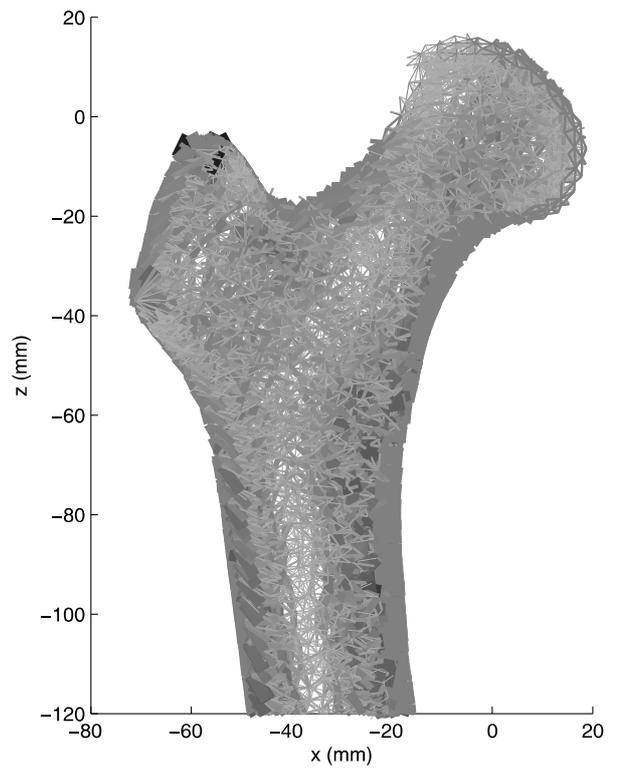
(b) 1st Iteration

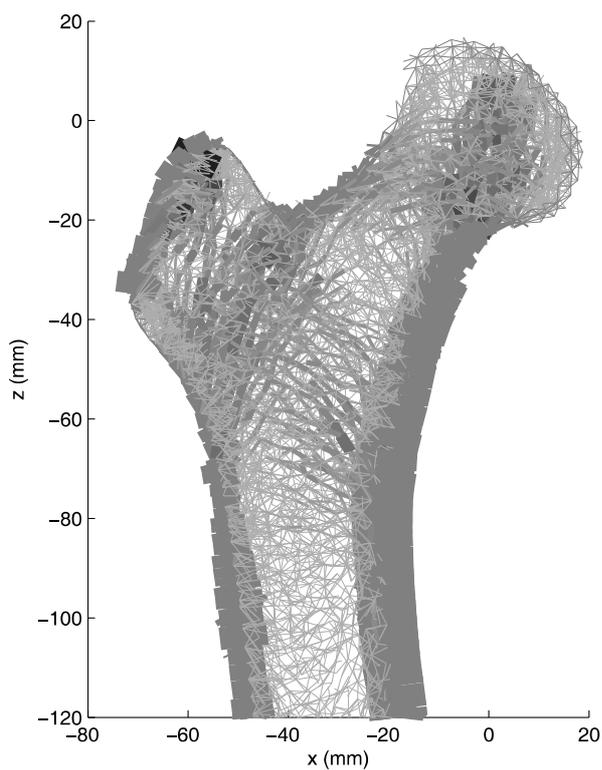
(c) 5th Iteration

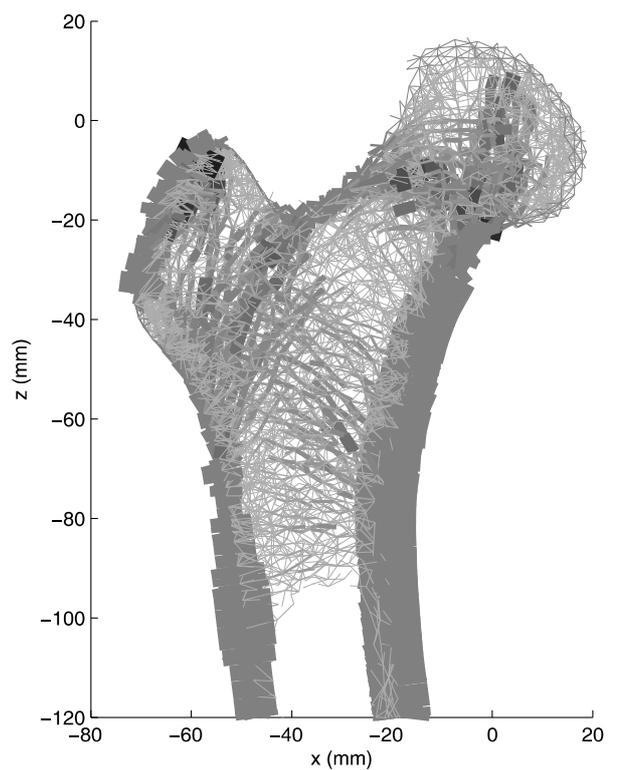
(d) Converged configuration

Figure 5: 5mm slice through the proximal femur for the (a) original, (b) 1st, (c) 5th and (d) converged iterations. Trabecular bone size and greyscale value is scaled based on the cross-sectional area; cortical bone size is scaled based on the thickness.

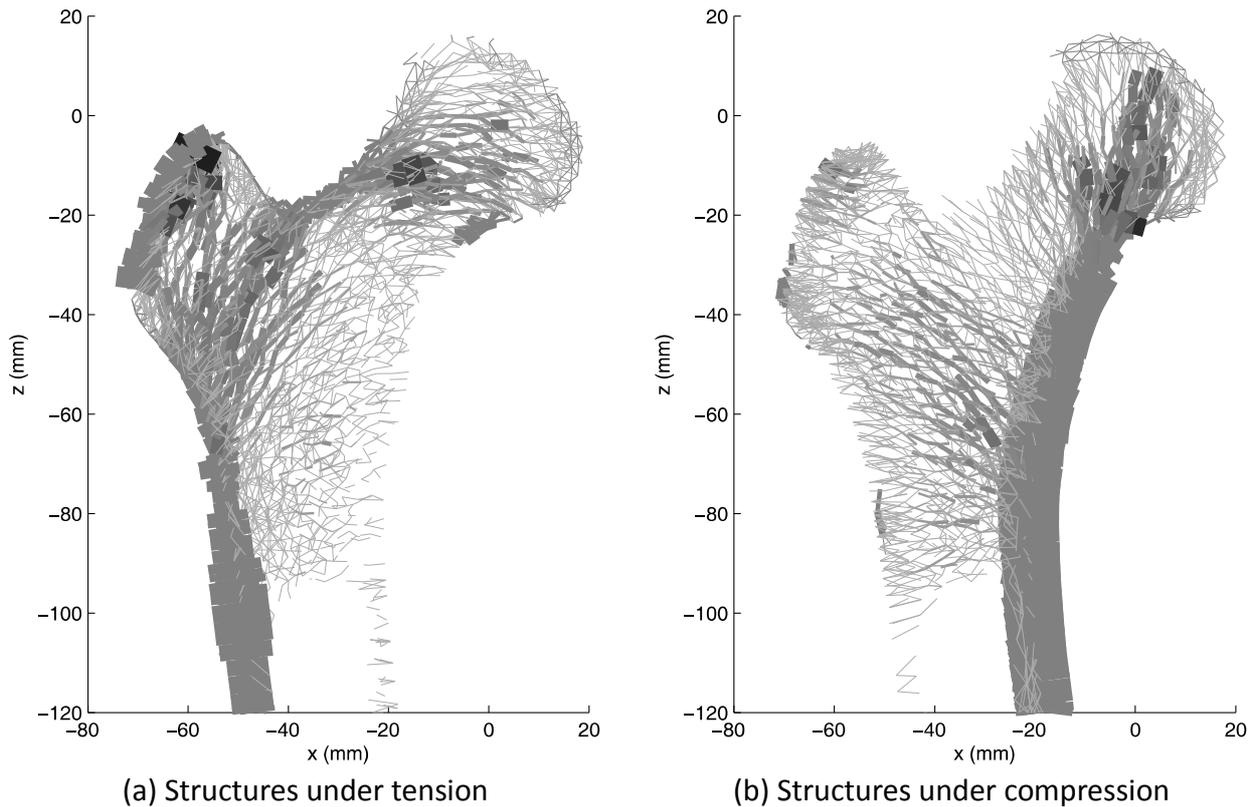

(a) Structures under tension     (b) Structures under compression

Figure 6: 5mm slice through the proximal femur for the converged state, showing (a) those structures predominantly under tension and (b) those structures predominantly under compression, based on the applied loading and boundary conditions.

**Discussion:**

From Figure 5 it is observed that following a massive stiffening of the femoral structure compared to the base model in the first iteration, convergence towards to final iteration is rapid, with the structure at the end of the fifth iteration being similar to the structure at the final iteration, with the exception of the presence of the femoral canal.

Comparing both the final iteration in Figure 5 and the structures under tension and compression in Figure 6 with the trabecular groups identified by Singh et al. (1970) a strong qualitative correlation is observed, with all five groups being present. In addition, Ward's triangle and the femoral canal are clearly visible. The thickness of the cortex, particularly thickening in the femoral shaft and a thin layer at the femoral head is consistent with previous clinical observations (Stephenson & Seedhom 1998, Treece et al. 2010).

For comparison with published values the bone volume to total volume ratio (BV/TV) in the femoral head was calculated based on a centralised sphere with a 10mm radius. This gave a value of 0.534, which correlates well with a measured maximum BV/TV in the femoral head of 0.481 reported by Hilderbrand et al. (1999). It was found however that the predicted value of BV/TV was dependent on the radius of the sphere chosen, increasing to 0.802 for a 5mm radius, and decreasing to 0.290 for a 15mm radius. Modelling limitations that would be expected to give this type of localisation include the use of a single load case, and the application of loads such as the hip joint contact force over a limited area.

It should be noted that despite the strong correlation between the predicted structure of the proximal femur found as part of this study, and previous clinical observations there are a number of limitations to the study. These include the use of axial truss elements for trabecular bone. This choice was made in order to restrict the choice of strain stimulus in the trabeculae to a single criteria. Physiologically it is clear that that trabeculae will be subject to further axial strain, as well as shear strain due to shearing, bending and twisting actions, in addition to direct axial deformation. Future versions of the model will investigate the use of other strain criteria in combination with beam elements. The modelling approach used in the study relied on using an initially high connectivity to produce a range of trabecular directions. While this was achieved it is clear that the resulting finite range of directions may not be sufficient to allow a fully optimised structure to develop. Future versions of the model will investigate methods of introducing trabeculae with optimised directions as part of the iterative modelling process.

Further limitations include the use of point loads to represent the joint contact and muscle loads, as well as the use of a reduced set of muscle forces, and simplified boundary conditions. In examining the proximal femur these assumptions are likely to lead to inaccuracy in the predicted structures around the greater trochanter and femoral head, while attempt has not been made as part of this study to investigate the distal femur, due to the use of fixed boundary conditions. In addition the study only considered a single load case, while it is clear that physiologically the structure of the femoral construct will be optimised for a range of activities. It is planned to address these limitations by combining the current iterative model with a soft or free boundary condition modelling approach developed previously (Phillips et al. 2007, Phillips 2009), over a range of activities to allow a more physiological representation of the action of muscles and ligaments surrounding the femur, as well as the use of distributed joint contact forces.

Both macro and micro-scale continuum models are seen to have distinct advantages and disadvantages. Macro-scale models are computationally efficient, but to date it has been found that they are limited to modelling bone material in an orthotropic manner. Micro-scale models represent the structure properties of bone, but are extremely computationally intensive. Tsubota et al. (2009) presented a detailed micro-scale iterative model of the femur using around 93 million hexahedral elements based on an 87.5µm voxel resolution. Results from this preliminary study are qualitatively similar using around 0.25% of the number of elements.

While appropriate modelling approaches should be adopted for particular situations, meso-scale structural modelling is seen as an attractive alternative that addresses some of the disadvantages of both macro and micro-scale continuum modelling.

**Conclusion:**

A structural optimisation approach has been used in assessing the biomechanics of the femur. Based on its similarity to clinical observations, the resulting predicted structure appears to support the concept of the proximal femur as a structure optimised to withstand the applied forces using a minimal amount of material. The use of a structural as opposed to a continuum approach allowed the creation of an extremely computationally efficient model, particularly in comparison to continuum models operating on the micro-scale. The developed trabecular structure was not limited to being orthotropic, in constrast to current macro-scale continuum models.

The use of a structural approach to modelling and understanding musculo-skeletal systems in biomechanics is seen to have potential as a complementary approach to other modelling techniques. It is intended that the approach developed in this study will be combined with other methods in order to produce a complete structural model of the musculo-skeletal system.

**Acknowledgements:**

The author would like to thank Nicholas Simpson and Dorothée Vallot, whose dissertation projects inspired the beginnings of the study, as well as the Structural Biomechanics Group for their thoughts and discussions on various aspects of the work presented here.

**References:**

Bergmann, G., Deuretzabacher, G., Heller, M., Graichen, F., Rohlmann, A., Strauss, J. and Duda, G., (2001). Hip forces and gait patterns from rountine activities, *Journal of Biomechanics*, **34**, 859-871.
Bitsakos, C., Kerner, J., Fisher, F. and Amis, A. A., (2005). The effect of muscle loading on the simulation of bone remodelling in the proximal femur, *Journal of Biomechanics*, **38**, 133-139.
Carter, D. R., Orr, T. E. and Fyhrie, D. P., (1989). Relationships between loading history and femoral cancellous bone architecture, *Journal of Biomechanics*, **22**, 231-244.
Culmann, K., *Die graphische Statik*. Verlag von Meyer & Zeller, 1866.
Frost, H. M., (2003). Bone's Mechanostat: A 2003 Update, *The Anatomical Record Part A*, **275A**, 1081-1101.
Geraldes, D. M. and Phillips, A. T. M., 3D strain-adaptive continuum orthotropic bone remodelling algorithm: prediction of bone architecture in the femur. *IFMBE Proceedings of the 6th World Congress of Biomechanics (WCB 2010)*, Singapore, 2010, **31**, 772-775.
Hilderbrand, T., Laib, A., Müller, R., Dequeker, J. and Rüegsegger, P., (1999). Direct three-dimensional morphometric analysis of human cancellous bone: microstructural data from spine, femur, iliac crest, and calcaneus, *Journal of Bone and Joint Surgery (Am)*, **14**, 1167-1174.
Koch, J. C., (1917). The laws of bone architecture, *The American Journal of Anatomy*, **21**, 177-298.
von Meyer, G., (1867). Die Architektur der Spongiosa, *Arch. Anat. Physiol. Wiss. Med.*, **34**, 615-628.
Miller, Z., Fuchs, M. B. and Arcan, M., (2002). Trabecular bone adaptation with an orthotropic material model, *Journal of Biomechanics*, **35**, 247-256.
Morlock, M., Schneider, E., Bluhm, A., Vollmer, M., Bergmann, G., Müller, V. and Honl, M., (2001). Duration and frequency of every day activities in total hip patients, *Journal of Biomechanics*, **34**, 873-881.
Phillips, A. T. M., (2009). The femur as a musculo-skeletal construct: a free boundary condition modelling approach, *Medical Engineering & Physics*, **31**, 673-670.
Phillips, A. T. M., Pankaj, P., Howie, C. R., Usmani, A. S. and Simpson, A. H. R. W., (2007). Finite element modelling of the pelvis: inclusion of muscular and ligamentous boundary conditions, *Medical Engineering & Physics*, **29**, 739-748.
Pidaparti, R. and Turner, C., (1997). Cancellous bone architecture: advantages of nonorthogonal trabecular alignment under multidirectional loading, *Journal of Biomechanics*, **30**, 979-983.
van Rietbergen, B., Huiskes, R., Eckstein, F. and Rüegsegger, P., (2003). Trabecular bone tissue strains in the healthy and osteoporotic human femur, *Journal of Bone and Mineral Research*, **18**, 1781-1788.
Singh, M., Nagrath, A. and Maini, P., (1970). Changes in trabecular pattern of the upper end of the femur as an index of osteoporosis, *Journal of Bone and Joint Surgery (Am)*, **52-A**, 457-467.
Skedros, J. G. and Baucom, S. L., (2007). Mathematical analysis of trabecular "trajectories" in apparent trajectorial structures: The unfortunate historical emphasis on the human proximal femur., *Journal of Theoretical Biology*, **244**, 15-45.
Stephenson, P. and Seedhom, B., (1998). Cross-sectional geometry of the human femur in the mid-third region, *Proceedings of the IMechE, Part H: Journal of Engineering in Medicine*, **213**, 159-166.

Treece, G. M., Gee, A. H., Mayhew, P. M. and Poole, K. E. S., (2010). High resolution cortical bone thickness measurement from clinical CT data, *Medical Image Analysis*, **14**, 276-290.

Tsubota, K.-i., Suzuki, Y., Yamada, T., Hojo, M., Makinouchi, A. and Adachi, T., (2009). Computer simulation of trabecular remodelling in human proximal femur using large-scale voxel FE models: approach to understanding Wolff's law, *Journal of Biomechanics*, **42**, 1088-1094.

Turner, C. H., Rho, J., Takano, Y., Tsui, T. Y. and Pharr, G. M., (1999). The elastic properties of trabecular and cortical bone tissues are similar: results from two microscopic measurement techniques, *Journal of Biomechanics*, **32**, 437-441.

Verhulp, E., van Rietbergen, B. and Huiskes, R., (2006). Comparison of micro-level and continuum-level voxel models of the proximal femur, *Journal of Biomechanics*, **39**, 2951-2957.

Wolff, J., *The Law of Bone Remodelling (original German edition title: Das Gesetz der Transformation der Knochen, 1892)*. Springer-Verlag, 1986.